\documentstyle[preprint,aps]{revtex}
\tighten
\draft
\widetext
\input epsf
\preprint{HUTP-98/A018, NUB 3174}
\bigskip
\bigskip
\begin{document}
\title{Gauge Theories from Orientifolds and Large $N$ Limit}
\medskip
\author{Zurab Kakushadze\footnote{E-mail: 
zurab@string.harvard.edu}}
\bigskip
\address{Lyman Laboratory of Physics, Harvard University, Cambridge, 
MA 02138\\
and\\
Department of Physics, Northeastern University, Boston, MA 02115}
\date{March 25, 1998}
\bigskip
\medskip
\maketitle

\begin{abstract}
{}Extending the recent work in \cite{BKV}, we consider 
string perturbative expansion in the presence of D-branes and orientifold
planes imbedded in orbifolded space-time. In the 
$\alpha^\prime\rightarrow 0$ limit the weak coupling string perturbative
expansion maps to `t Hooft's large $N$ expansion. We focus
on four dimensional ${\cal N}=1,2,4$ supersymmetric theories,
and also discuss possible extensions to ${\cal N}=0$ cases. Utilizing 
the string theory perturbation techniques we show that computation
of any $M$-point correlation function in these theories reduces to
the corresponding computation in the parent ${\cal N}=4$ theory. 
In particular, we discuss theories (which are rather constrained)
with vanishing $\beta$-functions
to all orders in perturbation theory in the large $N$ limit. We also
point out that in theories with non-vanishing $\beta$-functions the gauge
coupling running is suppressed in the large $N$ limit. 
Introduction of orientifold planes allows to construct certain gauge theories
with $SO$, $Sp$ and $SU$ gauge groups and various matter 
(only unitary gauge groups with bi-fundamental/adjoint
matter arise in theories without orientifold planes).

\end{abstract}

\section{Introduction}

{}`t Hooft's large $N$ expansion \cite{thooft} is an attractive possibility
for understanding gauge theories. In this limit the gauge theory diagrams
look like Riemann surfaces with boundaries and handles. It is therefore
natural to attempt to map the large $N$ expansion of gauge theories
to some kind of string world-sheet expansion.

{}The first concrete example of such a map was given by Witten \cite{CS} for 
the case of three dimensional Chern-Simons gauge theory where the 
boundaries of the string world-sheet are ``topological'' D-branes. 

{}Recently the map between the large $N$ expansion and string expansion
has been made precise \cite{BKV} in the context of Type II string theory. 
The idea of \cite{BKV} is to consider Type IIB string theory with a large number 
$N$ of D3-branes and take a limit where $\alpha^\prime\rightarrow 0$ while 
keeping $\lambda=N\lambda_s$ fixed, where $\lambda_s$ is the Type IIB string 
coupling. In this setup we have four dimensional 
gauge theories with unitary gauge groups. A 
world-sheet with $b$ boundaries and $g$ handles is weighted with
\begin{equation}\label{thoo}
 (N\lambda_s)^b \lambda^{2g-2}_s=\lambda^{2g-2+b} N^{-2g+2}~.
\end{equation} 
Upon identification $\lambda_s=g^2_{YM}$, this precisely maps to `t Hooft's
large $N$ expansion. This expansion is valid in the limit where $N\rightarrow\infty$
and the effective coupling $\lambda$ is fixed at a weak coupling value.

{}In \cite{BKV} the above idea was applied to prove that four dimensional 
gauge theories (including the cases with no supersymmetry) considered in
\cite{KaSi,LNV}\footnote{For other related works, see, {\em e.g.}, \cite{Iban,HSU}.} 
are conformal to all orders in perturbation theory in the large $N$ 
limit. In particular, the corresponding gauge theories were obtained from 
Type IIB string theory with D3-branes imbedded in orbifolded space-time.
The ultraviolet finiteness of string theory (that is, one-loop tadpole cancellation 
conditions) was shown to imply that the resulting (non-Abelian) gauge theories
where conformal in the large $N$ limit (in all loop orders). Moreover, in \cite{BKV}
it was also proven that computation of any correlation function in these theories 
in the large $N$ limit reduces to the corresponding computation in the parent 
${\cal N}=4$ supersymmetric gauge theory. 

{}The all-order proofs in \cite{BKV} were possible due to the fact that the power
of string perturbation techniques was utilized. In particular, string theory perturbation
expansion is a very efficient way of summing up various field theory diagrams.
Thus, often a large number of field theory Feynman diagrams in a given order of
perturbation theory corresponds to a single string theory diagram with certain
topology.
This has been successfully exploited to compute tree and loop level scattering 
processes in gauge theories \cite{Koso}. (For a recent discussion, see, {\em e.g.}, 
\cite{Dix}.) The all-order proofs in \cite{BKV} crucially depended on the fact that
the string world-sheet expansion was self-consistent. In particular, the arguments
in \cite{BKV} would not go through if the tadpoles were not cancelled. The tadpole
cancellation conditions, however, ultimately produced theories which were 
(super)conformal in the large $N$ limit. In particular, the one-loop $\beta$-functions 
(for non-Abelian gauge groups) in all of those theories were zero (even at finite $N$).

{}It is natural to ask whether there exists a consistent string theory setup similar to that 
of \cite{BKV} which can produce: ({\em i}) finite gauge theories with gauge groups
other than unitary (in particular, $SO$ and $Sp$), and with matter other than bi-fundamentals;
({\em ii}) non-finite gauge theories such that string theory perturbation techniques
can be applied to learn about the gauge theories in the large $N$ limit. In this paper we
consider precisely such a setup which is a generalization of the work in \cite{BKV}.
Here we will consider Type IIB string theory with D3-branes as well as orientifold planes.
(In certain cases string consistency will also require presence of D7-branes.)
That is, we are going to study gauge theories that arise from Type IIB orientifolds.
In this framework we can expect appearance of $SO$ and $Sp$ gauge groups as well
as matter other than bi-fundamentals (in the product gauge groups). Moreover, as we
will see later, non-finite gauge theories can also be obtained within a consistent 
perturbative string theory framework. However, in all such theories obtained via
orientifolds the running of the gauge couplings is suppressed in the large $N$ limit.
(In this sense these theories are ``finite'' in the large $N$ limit).

{}Introduction of orientifold planes changes the possible topologies of the world-sheet.
Now we can have a world-sheet with $b$ boundaries (corresponding to D-branes), 
$c$ cross-caps (corresponding to orientifold planes), and $g$ handles. Such a 
world-sheet is weighted with     
\begin{equation}\label{thoo1}
 (N\lambda_s)^b \lambda_s^c \lambda^{2g-2}_s=\lambda^{2g-2+b+c} N^{-c-2g+2}~.
\end{equation}
Note that addition of a cross-cap results in a diagram suppressed by an additional 
power of $N$, so that in the large $N$ limit we can hope for simplifications (or, rather,
we can hope to avoid complications with unoriented world-sheets, at least in some 
cases). In fact, we will show that for string vacua which are perturbatively consistent 
(that is, the tadpoles cancel) calculations of correlation functions 
in ${\cal N}<4$ gauge theories
reduce to the corresponding calculations in the parent ${\cal N}=4$
{\em oriented} theory. This holds not only for finite (in the large $N$ limit) gauge theories
but also for the gauge theories which are not conformal. We will prove that certain
${\cal N}=1$ gauge theories obtained this way are superconformal to all orders in 
perturbation theory in the large $N$ limit. We will also discuss possible extensions to 
non-supersymmetric cases.

{}It is very satisfying to observe that in the large $N$ limit
using the power of string theory perturbation techniques
we can reduce very non-trivial calculations in gauge theories with lower supersymmetries
to calculations in ${\cal N}=4$ gauge theories. In particular, this applies to
multi-point correlators in gauge theory.

{}Here we note that the correspondence between `t Hooft's large $N$ expansion and string 
world-sheet expansion is expected to hold only in the regime where the effective
coupling $\lambda$ is small. If $\lambda$ is large one expects an effective supergravity description to take over. This domain has been recently studied in several papers 
\cite{Kleb,Gubs,Mald,Poly,Oogu,Witt} as well as in various related works \cite{related}.
The supergravity picture has, in particular, led to the conjectures 
in \cite{KaSi} as well as in \cite{LNV} about 
finiteness of certain gauge theories. However, proofs of those conjectures 
(in the large $N$ limit)
presented in \cite{BKV} were given in the {\em weakly} coupled region. Also, $1/N$ corrections
can only be reliably computed in this region but not in the strong coupling regime where
{\em a priori} there is no world-sheet expansion nor `t Hooft's expansion is valid.

{}The remainder of this paper is organized as follows. In section II we review the
arguments of \cite{BKV} for the cases without the orientifold planes. In section III we
generalize
these arguments to the orientifold cases. 
In particular, we prove various vanishing theorems in the 
large $N$ limit, and give the relation between various correlators in ${\cal N}<4$ gauge 
theories to those in the parent ${\cal N}=4$ theory. In section IV we construct some explicit
${\cal N}=2$ examples. All of these examples turn out to be superconformal. In section V
we construct an example of ${\cal N}=1$ gauge theory with $Sp(N)$ gauge group. This 
gauge theory is superconformal in the large $N$ limit to all orders in perturbation theory.
We also discuss some examples of non-finite ${\cal N}=1$ gauge theories.
\section{Large $N$ Limit and Finiteness}\label{finite}

{}In this section we review the discussion in \cite{BKV} for the cases without orientifold
planes. We will generalize these arguments to the orientifold cases in section \ref{finite1}.

\subsection{Setup}

{}Consider Type IIB string theory with $N$ parallel D3-branes where
the space transverse to the
D-branes is ${\cal M}={\bf R}^6/\Gamma$.
The orbifold group
$\Gamma= \left\{ g_a \mid a=1,\dots,
|\Gamma| \right\}$ ($g_1=1$)
must be a finite discrete subgroup of $Spin(6)$.
If $\Gamma\subset SU(3)$ ($SU(2)$), we have
${\cal N}=1$ (${\cal N}=2$) unbroken supersymmetry,
and ${\cal N}=0$, otherwise.

{}Let us confine our attention to the cases where type IIB on ${\cal M}$ is
a modular invariant theory\footnote{This is always the case if $\Gamma\subset SU(3)$.
For the non-supersymmetric cases this is also true provided that
$\not\exists{\bf Z}_2\subset\Gamma$. If $\exists{\bf Z}_2\subset\Gamma$,
then modular invariance requires that the set of points in ${\bf R}^6$
fixed under the ${\bf Z}_2$ twist has real dimension 2.}. The action of the
orbifold on
the coordinates $X_i$ ($i=1,\dots,6$) on ${\cal M}$ can be described
in terms of $SO(6)$ matrices:
$g_a:X_i\rightarrow \sum_j (g_a)_{ij} X_j$.
We need to specify
the action of the orbifold group on the Chan-Paton charges carried by the
D3-branes. It is described by $N\times N$ matrices $\gamma_a$ that
form a representation of $\Gamma$. Note that $\gamma_1$ is an identity
matrix and ${\mbox {Tr}}(\gamma_1)=N$.

{}The D-brane sector of the theory is described by an {\it oriented} open
string theory. In particular, the world-sheet expansion corresponds
to summing over oriented Riemann surfaces with arbitrary genus $g$ and
arbitrary number of boundaries $b$, where the boundaries of the world-sheet 
correspond to the D3-branes.

{}For example, consider one-loop vacuum amplitude ($g=0$, $b=2$). The
corresponding graph is an annulus whose boundaries lie on D3-branes.
The one-loop partition function in the
light-cone gauge is given by
\begin{equation}\label{partition}
 {\cal Z}={1\over 2|\Gamma|}\sum_a
 {\rm Tr}  \left( g_a (1+(-1)^F)
 e^{-2\pi tL_0}
 \right)~,
\end{equation}
where $F$ is the fermion number operator, $t$ is the real modular parameter
of the annulus, and the trace includes sum over the Chan-Paton factors.

{}The orbifold group $\Gamma$ acts on both ends of the open strings.
The action of $g_a\in \Gamma$ on Chan-Paton charges is given by
$\gamma_a\otimes \gamma_a$. Therefore,
the individual terms in the sum in (\ref{partition})
have the following form:
\begin{equation}
 \left({\mbox {Tr}}(\gamma_a)\right)^2 {\cal Z}_a~,
\end{equation}
where ${\cal Z}_a$ are characters
corresponding to the world-sheet degrees of freedom. The ``untwisted''
character
${\cal Z}_1$ is the same as in the ${\cal N}=4$ theory for which
$\Gamma=\{1\}$. The information about the fact that the orbifold theory
has reduced supersymmetry is encoded in the ``twisted'' characters
${\cal Z}_a$, $a\not=1$.

{}In \cite{BKV} it was shown that the one-loop massless (and, in non-supersymmetric 
cases, tachyonic) tadpole cancellation conditions require that
\begin{equation}\label{pole}
 {\mbox {Tr}}(\gamma_a)=0~\forall a\not=1~.
\end{equation}
It was also shown that this condition implies that the Chan-Paton matrices $\gamma_a$
form an $n$-fold copy of the {\em regular} representation of $\Gamma$. The regular representation decomposes into a direct sum of all irreducible
representations ${\bf r}_i$ of $\Gamma$ with degeneracy factors
$n_i=|{\bf r}_i|$. The gauge group is ($N_i\equiv nn_i$)
\begin{equation}
 G=\otimes_i U(N_i)~. 
\end{equation}
The matter consists of Weyl fermions (and scalars) transforming in
bi-fundamentals
$({\bf N}_i,{\overline {\bf N}}_j)$ according to the decomposition
of the tensor product
of ${\bf 4}$ (${\bf 6}$) of $Spin(6)$ with the corresponding representation
(see \cite{LNV} for details).

\subsection{Large $N$ Limit}

{}The gauge group in the theories we are considering here
is $G=\otimes_i U(N_i) (\subset  U(N))$. In the following we will ignore the
$U(1)$ factors (for which the gauge couplings do run for ${\cal N}<4$ as there
are matter fields charged under them) and consider
$G=\otimes_i SU(N_i)$. In this subsection we review the arguments of
\cite{BKV} which show that in the large $N$
limit this non-Abelian gauge theory is 
conformal\footnote{Including the $U(1)$ factors does not alter the conclusions 
as their effect is subleading in the $1/N$ expansion \cite{BKV}.}.

{}There are two classes of diagrams we need to consider: ({\it i}) diagrams
without handles;
({\it ii}) diagrams with handles. The latter correspond to closed string loops
and are
subleading in the large $N$ limit. The diagrams without handles can 
be divided into two classes: ({\it i}) planar
diagrams
where all the external lines are attached to the same boundary; ({\it ii})
non-planar diagrams
where the external lines are attached to at least two different boundaries. The
latter are subleading in the large $N$ limit.

{}In the case of planar diagrams we have $b$ boundaries with all $M$ 
external lines attached to the same boundary (which without loss of generality
can be chosen to be the outer boundary) as depicted in Fig.1.
We need to sum over all possible twisted boundary conditions for the boundaries.
The boundary conditions must satisfy the requirement that
\begin{equation}\label{mono}
 \gamma_{a_1}=\prod_{s=2}^{b} \gamma_{a_s}~,
\end{equation}
where $\gamma_{a_1}$ corresponds to the outer boundary (to which we have attached
the external lines), and 
$\gamma_{a_s}$ ($s=2,\dots,b$) correspond to the inner boundaries (with no external lines).
Here we have chosen the convention (consistent with the corresponding choice made
for the annulus amplitude in (\ref{partition})) that the outer and inner boundaries have opposite
orientations. Then the above condition is simply the statement that only the states
invariant under the action of the orbifold group contribute into the amplitude.

{}If all the twisted boundary conditions are trivial ({\em i.e.}, $a_s=1$ for all $s=1,\dots,b$)
then the corresponding amplitude is the same as in the 
${\cal N} =4$ case (modulo factors of $1/\sqrt{|\Gamma|}$ coming from 
the difference in normalizations of the corresponding D-brane boundary states
in the cases with ${\cal N}=4$ (where $|\Gamma|=1$) and ${\cal N}<4$ (where 
$|\Gamma|\not=1$)). Therefore, such amplitudes do not contribute to the gauge
coupling running (for which we would have $M=2$ gauge bosons attached to
the outer boundary) since the latter is not renormalized in ${\cal N}=4$ gauge 
theories due to supersymmetry. 

{}Let us now consider contributions with non-trivial twisted
boundary conditions. Let $\lambda_r$, $r=1\dots M$, be the Chan-Paton matrices corresponding to the
external lines. Then the planar diagram with $b$ boundaries has the following
Chan-Paton group-theoretic dependence:
\begin{equation}
 \sum {\rm Tr}\left(\gamma_{a_1} \lambda_1\dots\lambda_M\right)
\prod_{s=2}^{b}  {\rm Tr}(\gamma_{a_s})~,
\end{equation}
where the sum involves all possible distributions of $\gamma_{a_s}$ twists
(that satisfy the condition (\ref{mono})) as well as permutations of $\lambda_r$
factors (note that the $\lambda$'s here are the states
which are kept after the orbifold projection, and so they commute 
with the action of $\gamma$'s). The important point here
is that unless all twists $\gamma_{a_s}$ are trivial for $s=2,\dots,b$, the
above diagram vanishes by the virtue of
(\ref{pole}). But then from (\ref{mono}) it follows that $\gamma_{a_1}$
must be trivial as well. This implies that the only planar diagrams that
contribute are those with trivial boundary conditions which (up to numerical
factors) are the same as in the parent ${\cal N}=4$ gauge theory. 
This establishes that computation of any $M$-point correlation faction in the
large $N$ limit reduces to the corresponding ${\cal N}=4$ calculation, and that
these gauge theories are (super)conformal in this limit\footnote{In \cite{BKV} it was also 
shown that a large class of non-planar diagrams without handles also vanish. We refer
the reader to \cite{BKV} for details.}.

{}Here we should mention that the models of \cite{BKV} are perturbatively
consistent string theories at all energy scales. In particular, the Abelian factors (that 
run in the low energy effective theory and decouple in the infrared) are not problematic
from the string theory viewpoint (although in the field theory context they would have
Landau poles in the ultraviolet).

\section{Generalization to Orientifolds}\label{finite1}

{}In this section we generalize the approach of \cite{BKV} to theories with
orientifold planes. Here we will mostly concentrate on supersymmetric cases, and
briefly discuss possible generalizations to non-supersymmetric cases
at the end of this section.  

\subsection{Setup}

{}Consider Type IIB string theory on ${\cal M}={\bf C}^3/\Gamma$ where
$\Gamma\subset SU(3)$ so that the resulting theory has some number
of unbroken supersymmetries. Consider the $\Omega J$ orientifold of this 
theory, where $\Omega$ is the world-sheet parity reversal, and $J$ 
is a ${\bf Z}_2$ element ($J^2=1$) acting on the complex coordinates $z_i$
($i=1,2,3$) on ${\bf C}^3$ such that the set of points in ${\bf C}^3$ fixed under 
the action of $J$ has real dimension $\Delta=0$ or $4$. 

{}If $\Delta=0$ then we have an orientifold 3-plane. If $\Gamma$ has
a ${\bf Z}_2$ subgroup, then we also have an orientifold 7-plane.
If $\Delta=4$ then we have an orientifold 7-plane. We may also have
an orientifold 3-plane depending on whether $\Gamma$ has an appropriate
${\bf Z}_2$ subgroup. Regardless of whether we have an orientifold 3-plane,
we can {\em a priori} introduce an arbitrary number of D3-branes (as the corresponding
tadpoles vanish due to the fact that the space transverse to the D3-branes
is non-compact). On the other hand, if we have an orientifold 7-plane we must 
introduce 8 of the corresponding D7-branes to cancel the R-R charge appropriately.
(The number 8 of D7-branes is required by the corresponding tadpole cancellation
conditions.) 

{}We need to specify the action of $\Gamma$ on the Chan-Paton factors
corresponding to the D3- and/or D7-branes. Just as in the previous section, 
these are given by Chan-Paton matrices which we collectively refer to
as $\gamma^\mu_a$, where the superscript $\mu$ refers to the corresponding
D3- or D7-branes. Note that ${\mbox{Tr}}(\gamma^\mu_1)=n^\mu$ where 
$n^\mu$ is the number of D-branes labelled by $\mu$. 

{} At one-loop level there are three different sources for massless tadpoles:
the Klein bottle, annulus, and M{\"o}bius strip amplitudes depicted in Fig.2. 
The Klein bottle amplitude corresponds to the contribution of unoriented 
closed strings into one-loop vacuum diagram. It can be alternatively viewed
as a tree-level closed string amplitude where the closed strings propagate 
between two cross-caps. The latter are (coherent Type IIB) states 
that describe the familiar orientifold planes. The annulus amplitude 
corresponds to the contribution of open strings stretched between two D-branes
into one-loop vacuum amplitude. It can also be viewed as a tree-channel closed
string amplitude where the closed strings propagate between two D-branes.
Finally, the M{\"o}bius strip amplitude corresponds to the contribution of unoriented
open strings into one-loop vacuum diagram. It can be viewed as a tree-channel closed 
string amplitude where the closed strings propagate between a D-brane and an orientifold
plane.

{}Note that there are no Chan-Paton matrices associated with the Klein bottle 
amplitude since it corresponds to closed strings propagating between two cross-caps
which do not carry Chan-Paton charges. The M{\"o}bius strip has only one boundary.
This implies that the individual terms (corresponding to twists $g_a\in \Gamma$)
in the M{\"o}bius strip amplitude are proportional to ${\mbox{Tr}}(\gamma^\mu_a)$. The annulus
amplitude is the same (up to an overall factor of $1/2$ due to the orientation reversal projection) 
as in the oriented case discussed in the previous section. Thus, the individual terms (corresponding to twists $g_a\in \Gamma$)
in the annulus amplitude are proportional to ${\mbox{Tr}}(\gamma^\mu_a){\mbox{Tr}}(\gamma^\nu_a)$. 
Thus, the tadpoles can be written as
\begin{equation}\label{KMA}
 \sum_a \left( K_a +\sum_\mu M^\mu_a {\mbox{Tr}}(\gamma^\mu_a)+
                         \sum_{\mu,\nu}A^{\mu\nu}_a
 {\mbox{Tr}}(\gamma^\mu_a){\mbox{Tr}}(\gamma^\nu_a)
 \right)~.
\end{equation}   
Here terms with $K_a$, $M^\mu_a$ and $A^{\mu\nu}_a$ correspond to the
contributions of the Klein bottle, M{\"o}bius strip and annulus amplitudes, respectively.
In fact, the factorization property of string theory implies that the Klein bottle amplitude
should factorize into two cross-caps connected via a long thin tube. The M{\"o}bius strip
amplitude should factorize into a cross-cap and a disc connected via a long thin tube.
Similarly, the annulus amplitude should factorize into two discs connected via a long thin tube.
These factorizations are depicted in Fig.3. The implication of this for the tadpoles is that
they too factorize into a sum of perfect squares
\begin{equation}\label{tad}
 \sum_a \left(B_a+\sum_\mu C^\mu_a {\mbox{Tr}}(\gamma^\mu_a)\right)^2~,
\end{equation}  
where $B_a^2=K_a$, $2B_a C^\mu_a=M^\mu_a$ and $C^\mu_a C^\nu_a=A^{\mu\nu}_a$.
Thus, the tadpole cancellation conditions read:
\begin{equation}
 B_a+\sum_\mu C^\mu_a {\mbox{Tr}}(\gamma^\mu_a)=0~.
\end{equation}
Note that 
\begin{equation}\label{Klein}
 {\mbox {Tr}}(\gamma^\mu_a)=0~\forall a\not=1~{\mbox{only if}}~K_a=0~\forall a\not=1~.
\end{equation}
In the next subsection we will see that if this condition is satisfied then the
corresponding (non-Abelian) gauge theories are superconformal in the large $N$ limit.
On the other hand, if not all $K_a$ with $a\not=1$ are zero, then some of the Chan-Paton
matrices $\gamma^\mu_a$ with $a\not=1$ must have non-zero traces. This will
generically lead to theories with non-vanishing one-loop $\beta$-functions.

\subsection{Large $N$ Limit}

{}In this subsection we extend the arguments reviewed in section \ref{finite}
to the cases with orientifold planes. In particular, we will study the large $N$ behavior
of $M$-point correlators of fields charged under the gauge group that arises
from the D3-branes. In the following we will ignore the
$U(1)$ factors (if any) in the D3-brane gauge group.

{}There are two classes of diagrams we need to consider: ({\it i}) diagrams
without handles and cross-caps;
({\it ii}) diagrams with handles and/or cross-caps. 
The latter are
subleading in the large $N$ limit. The diagrams without handles and cross-caps
can 
be divided into two classes: ({\it i}) planar
diagrams
where all the external lines are attached to the same boundary; ({\it ii})
non-planar diagrams
where the external lines are attached to at least two different boundaries. The
latter are subleading in the large $N$ limit.

{}In the case of planar diagrams we have $b$ boundaries
corresponding to D3- and/or D7-branes. We will attach $M$ 
external lines to the outer boundary as depicted in Fig.1.
We need to sum over all possible twisted boundary conditions for the boundaries.
The boundary conditions must satisfy the requirement that
\begin{equation}\label{mono1}
 \gamma^{\mu_1}_{a_1}=\prod_{s=2}^{b} \gamma^{\mu_s}_{a_s}~,
\end{equation}
where $\gamma^{\mu_1}_{a_1}$ corresponds to the outer boundary, and 
$\gamma^{\mu_s}_{a_s}$ ($s=2,\dots,b$) 
correspond to the inner boundaries.

{}Let us first consider the ${\cal N}=4$ theories for which the orbifold group
$\Gamma$ is trivial. (Note that in this case we can only have D3-branes
as introduction of D7-branes would break some number of supersymmetries.) 
The computation of any correlation function in the orientifold
theory (with $SO(N)$ or $Sp(N)$ gauge group) is reduced to the corresponding
computation in the oriented ${\cal N}=4$ theory (with $U(N)$ gauge group) up to factors
of $1/\sqrt{2}$ (coming from the difference in normalizations of the corresponding D-brane boundary states in the oriented and unoriented cases). Such a simplification is due to
the fact that the unoriented world-sheets with cross-caps give contributions suppressed
by extra powers of $1/N$. 

{}Next, consider cases where $\Gamma$ is non-trivial (and hence supersymmetry is reduced).
If all the twisted boundary conditions are trivial ({\em i.e.}, $a_s=1$ for all $s=1,\dots,b$)
then the corresponding amplitude is the same as in the 
${\cal N} =4$ case (modulo factors of $1/\sqrt{|\Gamma|}$).
Therefore, such amplitudes do not contribute to the gauge
coupling running (for which we would have $M=2$ gauge bosons attached to
the outer boundary) since the latter is not renormalized in ${\cal N}=4$ gauge 
theories due to supersymmetry. 

{}Let us now consider contributions with non-trivial twisted
boundary conditions. Let $\lambda_r$, $r=1\dots M$, be the Chan-Paton matrices corresponding to the
external lines. Then the planar diagram with $b$ boundaries has the following
Chan-Paton group-theoretic dependence:
\begin{equation}
 \sum {\rm Tr}\left(\gamma^{\mu_1}_{a_1} \lambda_1\dots\lambda_M\right)
\prod_{s=2}^{b}  {\rm Tr}(\gamma^{\mu_s}_{a_s})~,
\end{equation}
where the sum involves all possible distributions of $\gamma_{a_s}$ twists
(that satisfy the condition (\ref{mono})) as well as permutations of $\lambda_r$
factors. If the condition (\ref{Klein}) is satisfied, {\em i.e.}, if all the twisted Chan-Paton 
matrices are traceless, then the situation is analogous to that in the
oriented cases. That is, the only planar diagrams that
contribute are those with trivial boundary conditions. 
Such diagrams with all the boundaries corresponding to D3-branes (up to numerical
factors) are the same as in the parent ${\cal N}=4$ gauge theory.
The diagrams with trivial boundary conditions but with some boundaries corresponding
to D7-branes are subleading in the large $N$ limit as the numbers of D7-branes
(that is, the traces
${\mbox{Tr}}(\gamma^\mu_1)$ corresponding to D7-branes) are of order one.
This establishes that computation of any $M$-point correlation faction in the
large $N$ limit reduces to the corresponding ${\cal N}=4$ calculation in {\em oriented} 
theory, and that
these gauge theories are superconformal in this limit\footnote{Just as in
\cite{BKV}, it is also straightforward to show that 
a large class of non-planar diagrams without handles and cross-caps also vanish.}.

{}Now consider the cases where some of the twisted Chan-Paton matrices are {\em not}
traceless. Then there are going to be corrections to the $M$-point correlators coming
from planar diagrams with non-trivial twisted boundary conditions. These diagrams are 
subleading in the large $N$ limit as the corresponding traces are always of order one. 
This follows from the tadpole cancellation conditions (\ref{tad}) where the coefficients
$B_a$ and $C^\mu_a$ are of order one, so for $a\not=1$ we have ${\mbox{Tr}}(\gamma^\mu_a)\sim 1$. This implies that even for non-finite theories
computation of the correlation functions reduces to the corresponding computation in the
parent ${\cal N}=4$ oriented theory. 

{}Here we should point out that ``non-finiteness'' of such theories is a subleading 
effect in the large $N$ limit. This is because the $\beta$-function coefficients grow
as
\begin{equation}\label{beta}
 b_s=O(N^s)~,~~~s=0,1,\dots~,
\end{equation}
instead of $b_s=O(N^{s+1})$ (as in, say, pure $SU(N)$ gauge theory). This can be seen
by considering planar diagrams with $M=2$ external lines corresponding to gauge bosons.
Note that in string theory running of the gauge couplings in the low energy effective
field theory is due to infrared divergences corresponding to massless modes \cite{kap}.
The diagrams with all the boundaries corresponding to D3-branes and with all the 
boundary conditions corresponding to the identity element of $\Gamma$ are the same
(up to overall numerical factors) as in the parent ${\cal N}=4$ theory. Such diagrams,
therefore, do not contain infrared divergencies, and thus do not contribute the gauge 
coupling running. (That is, their contributions to the $\beta$-function coefficients $b_s$
vanish.) Therefore, the only diagrams that can contribute to the $\beta$-function coefficients $b_s$ are those with some boundaries corresponding to D7-branes and/or having twisted boundary
conditions with ${\mbox{Tr}}(\gamma^\mu_a)\not=0$. These are, however,
suppressed at least by one power of $N$ since the numbers of D7-branes are of order one,
and also such ${\mbox{Tr}}(\gamma^\mu_a)\sim 1$. This establishes (\ref{beta}).

{}Note that the estimates for the $\beta$-function coefficients in (\ref{beta}) for $b_{s>0}$
are non-trivial from the field theory point of view as they imply infinitely many cancellations
between couplings (such as Yukawas) in the gauge theory. On the other hand, within 
string perturbation expansion these statements become obvious once we carefully 
consider twisted boundary conditions and tadpole cancellation.

\subsection{Comments}

{}Here we would like to comment on some issues concerning the discussion 
in the previous subsections.

{}First, note that the entire argument in the previous subsection crucially depends
on the assumption that there is a well defined world-sheet description of the
orientifold theories at hand. Naively, it might seem that orientifolds of Type IIB
on ${\bf C}^3/\Gamma$ should have such world-sheet descriptions for any
$\Gamma$ which is a subgroup of $Spin(6)$. This is, however, not the case 
\cite{KST}\footnote{Some examples of this were also given in \cite{Zwart}.}.
In particular, there are certain cases where perturbative description is
inadequate as there are additional states present in the corresponding orientifolds
such that they are non-perturbative from the orientifold viewpoint \cite{KST}.
In section \ref{N1} we will give an explicit example of this. For further details we refer the reader
to \cite{KST}. 

{}The second comment is on possible extensions of the above results to ${\cal N}=0$
theories. Note that in the cases without orientifold planes \cite{BKV} the tadpole 
cancellation conditions for both ${\cal N}=0$ and supersymmetric theories are given 
by (\ref{pole}) for which there always exists a solution for any choice of $\Gamma$.
The situation in the cases with orientifold planes might not be so simple: the tadpoles
are of the form (\ref{KMA}), and the coefficients $K_a,M^\mu_a,A^{\mu\nu}_a$
may be different for the massless and tachyonic tadpoles. (The latter are absent in
the supersymmetric cases but are generically present for ${\cal N}=0$.) It is {\em a priori}
unclear whether the massless and tachyonic tadpole cancellation conditions are 
compatible. However, if they have a solution in a given model, then the statements
about the correlation functions in subsection B persist for such ${\cal N}=0$ models
in the large $N$ limit. It would therefore be interesting to construct ${\cal N}=0$
models in which all the tadpoles cancel.

\section{${\cal N}=2$ Superconformal Gauge Theories} 

{}In this section we construct four dimensional ${\cal N}=2$ superconformal gauge 
theories form Type IIB orientifolds. We start from Type IIB string theory on ${\cal M}$=
${\bf C}^3/\Gamma$, where $\Gamma=\{g^k\vert k=0,\dots,M-1\}\approx {\bf Z}_M$ 
is the orbifold group whose action on the complex coordinates $z_i$ ($i=1,2,3$) 
on ${\bf C}^3$ is given by $gz_1=z_1$, $gz_2=\omega z_2$, $gz_3=\omega^{-1} z_3$ 
($\omega=\exp(2\pi i/M)$). Next, we consider an orientifold of this theory where the 
orientifold action is given by $\Omega J$. Here $\Omega$ is the world-sheet parity 
reversal, and $J$ acts on $z_i$ as $Jz_i=-z_i$. The orientifold group is given by
${\cal O}=\{g^k,\Omega J g^k\vert k=0,\dots,M-1\}$. 

\subsection{Orbifolds of Even Order} 

{}Let us first consider the cases where $M$ is even. Then the $\Omega J$ 
orientifold of Type IIB on ${\cal M}$ is equivalent to the $\Omega {\widetilde J}$
orientifold of Type IIB on ${\cal M}$, where ${\widetilde J}$ acts on $z_i$ as
${\widetilde J}z_1=-z_1$, ${\widetilde J}z_2=z_2$, ${\widetilde J}z_3=z_3$.
(This is due to the presence of the ${\bf Z}_2$ element $R$ in $\Gamma$ where 
$Rz_1=z_1$, $Rz_2=-z_2$, $Rz_3=-z_3$. Note that ${\widetilde J}=JR$.) These 
orientifolds are similar to those considered in \cite{PS,GP,GJ}. In particular, 
calculation of massless tadpoles closely parallels that of \cite{GJ}. 

{}The orientifolds (with $M\in 2{\bf N}$) we are considering here contain 
both D3- and D7-branes. The action of the orbifold group on the Chan-Paton 
charges carried by the D3- and D7-branes is described by matrices 
$\gamma_{k,3}$ and $\gamma_{k,7}$ that form a (projective) 
representation\footnote{{\em A priori} they can form a representation of the double
cover of $\Gamma$.} of $\Gamma$. Note that $\gamma_{0,3}$ and $\gamma_{0,7}$
are identity matrices, and ${\mbox{Tr}}(\gamma_{0,3})=n_3$ and  ${\mbox{Tr}}(\gamma_{0,7})=n_7$, where $n_3$ and $n_7$ are the numbers of D3- 
and D7-branes, respectively.
 
{}The massless tadpoles can be computed following \cite{GJ}.
There are two types of massless tadpoles we need to consider. The first type of tadpoles correspond to untwisted closed string exchange in the tree-channel. Cancellation of 
untwisted tadpoles implies that there are $n_7=8$ D7-branes present (for even $M$). 
On the other hand, the number of D3-branes is unconstrained by the untwisted tadpole 
cancellation conditions. This is due to the fact that the space transverse to the D3-branes
is non-compact in the theories we are considering here.
 
{}Next, consider the tadpoles corresponding to twisted closed string exchange in the 
tree-channel. For even $M$ the
twisted massless tadpole cancellation conditions read:
\begin{eqnarray}
 &&{\mbox{Tr}}(\gamma_{2k-1,7})-4\sin^2\left({(2k-1)\pi\over M}\right)
 {\mbox{Tr}}(\gamma_{2k-1,3})=0~,\\
 &&{\mbox{Tr}}(\gamma_{2k,7})-4\sin^2\left({2\pi k\over M}\right)
 {\mbox{Tr}}(\gamma_{2k,3})-8\cos\left({2\pi k\over M}\right)=0~.
\end{eqnarray}
Note that the difference between these tadpole cancellation conditions and those
of \cite{GJ} is that in the second line we have $8\cos(2\pi k/M)$ instead of
 $32\cos(2\pi k/M)$. This is due to the fact that we are considering a system of D3- and 
D7-branes instead of a system of D5- and D9-branes\footnote{The former can be 
obtained from 
the latter by compactifying on $T^2$, T-dualizing, and taking the dimensions 
of the dual torus ${\widetilde T}^2$ to infinity. 
The orientifold $p$-planes ($p=5,9$) split into 4 orientifold 
$(p-2)$-planes upon T-dualizing. These are located at four fixed points of 
${\widetilde T}^2/{\bf Z}_2$. However, after taking the dimensions of 
${\widetilde T}^2$ to infinity ({\em i.e.}, when considering ${\bf C}/{\bf Z}_2$ instead of
${\widetilde T}^2/{\bf Z}_2$), only one fixed point (at the origin) remains. This results in 
reduction of all the tadpoles by a factor of 4. In particular, the number of D7-branes is 8 as opposed to 32 in the case of D9-branes.}.

{}It is not difficult to show that the above tadpole cancellation conditions have 
solutions\footnote{Here we must take into account that the Chan-Paton matrices 
$\gamma_{k,3}$ and $\gamma_{k,7}$ must form a (projective) representation of 
the orbifold group $\Gamma$ (or its double cover). Moreover, the orientifold 
projection $\Omega$ for the D-branes must be such that before taking into 
account the orbifold projections the gauge groups coming from the D3- and 
D7-branes must be $Sp$ and $SO$, respectively \cite{PS,GP}.} only for $M=2,4,6$.
The corresponding Chan-Paton matrices (up to equivalent representations) 
are given by:\\
$\bullet$ $M=2$ ($N=n_3/2$):
\begin{eqnarray}
 &&\gamma_{1,3}={\mbox{diag}}(i~(N~{\mbox{times}}), 
 -i~(N~{\mbox{times}}))~,\\
 &&\gamma_{1,7}={\mbox{diag}}(i~(4~{\mbox{times}}), -i~(4~{\mbox{times}}))~.
\end{eqnarray}
$\bullet$ $M=4$ ($N=n_3/4$):
\begin{eqnarray}
 \gamma_{1,3}={\mbox{diag}}(&&\exp(\pi i/4)~(N~{\mbox{times}}),
 \exp(-\pi i/4)~(N~{\mbox{times}}),\nonumber\\
 &&\exp(3\pi i/4)~(N~{\mbox{times}}),
 \exp(-3\pi i/4)~(N~{\mbox{times}}))~,\\
 \gamma_{1,7}={\mbox{diag}}(&&\exp(\pi i/4)~(2~{\mbox{times}}),
 \exp(-\pi i/4)~(2~{\mbox{times}}),\nonumber\\
 &&\exp(3\pi i/4)~(2~{\mbox{times}}),
 \exp(-3\pi i/4)~(2~{\mbox{times}}))~.
\end{eqnarray}
$\bullet$ $M=6$ ($N=(n_3-2)/6$):
\begin{eqnarray}
 \gamma_{1,3}={\mbox{diag}}(&&
 i\exp(2\pi i/3)~(N~{\mbox{times}}),
 -i\exp(2\pi i/3)~(N~{\mbox{times}}),\nonumber\\
 &&i\exp(-2\pi i/3)~(N~{\mbox{times}}),
 -i\exp(-2\pi i/3)~(N~{\mbox{times}})),\nonumber\\
 &&i~((N+1)~{\mbox{times}}),
 -i~((N+1)~{\mbox{times}}))~,\\ 
 \gamma_{1,7}={\mbox{diag}}(&&
 i\exp(2\pi i/3),-i\exp(2\pi i/3),i\exp(-2\pi i/3),-i\exp(-2\pi i/3),\nonumber\\
 &&i~(2~{\mbox{times}}), -i~(2~{\mbox{times}}))~.
\end{eqnarray}
The massless spectra (including twisted closed string sectors) of these models
are given in Table \ref{spectrum1}.

{}Note that the one-loop $\beta$-functions for the non-Abelian subgroups of the 33
open string sector gauge groups vanish in these models\footnote{The Abelian gauge 
couplings run in these models so that $U(1)$'s decouple at low energies.}. Thus, we 
have ${\cal N}=2$ superconformal gauge theories living in the world-volumes
of the D3-branes\footnote{The 77 gauge groups for large enough values of $N$ are 
infrared free. They can therefore be treated as global symmetries (in the context
of four dimensional gauge theories living on the D3-branes) at low energies.}.

\subsection{Orbifolds of Odd Order}

{}Let us now discuss the $\Omega J$ orientifold of Type IIB on ${\cal M}=
{\bf C}^3/\Gamma$, where the orbifold group $\Gamma\approx{\bf Z}_M$ has
odd order ($M\in2{\bf N}+1$). Such an orientifold contains an arbitrary number
of D3-branes. The number of D7-branes, however, is now zero which follows from
the untwisted tadpole cancellation conditions.  

{}To understand the twisted tadpole cancellation conditions, first consider 
$\Omega {\widetilde J}$ orientifold of  Type IIB on ${\cal M}$ (for odd $M$). 
This theory has only D7-branes and no D3-branes. The twisted tadpole 
cancellation conditions for the $\Omega J$ orientifold of Type IIB on ${\cal M}$
are isomorphic (after interchanging the corresponding D3- and D7-brane
Chan-Paton matrices) to those for the $\Omega {\widetilde J}$ orientifold of  
Type IIB on ${\cal M}$ provided that the tadpole cancellation conditions
for the $\Omega {\widetilde J}$ orientifold of Type IIB on ${\cal M}^\prime=
{\bf C}^3/\Gamma^\prime$ can be satisfied, where $\Gamma^\prime=\{g^k,Rg^k
\vert k=0,\dots,M-1\}\approx{\bf Z}_{2M}$. If the latter condition is not satisfied, then
there is no solution to the twisted tadpole cancellation conditions for the 
 $\Omega J$ orientifold of Type IIB on ${\cal M}$. The discussion in the previous 
subsection then implies that the twisted tadpole cancellation conditions can be
satisfied only for the $\Omega J$ orientifold of Type IIB on ${\cal M}$ with $M=3$.

{}The twisted tadpole cancellation conditions for the $\Omega {\widetilde J}$ 
orientifold of  Type IIB on ${\cal M}$ (for arbitrary odd $M$) can be computed
following \cite{GJ}, and are given by: 
\begin{eqnarray}\label{odd}
 {\mbox{Tr}}(\gamma_{2k,7})-4\sin^2\left({2\pi k\over M}\right)
 {\mbox{Tr}}(\gamma_{2k,3})-8\cos^2\left({\pi k\over M}\right)=0~.
\end{eqnarray}
(Note that the difference between these tadpole cancellation conditions and those 
of \cite{GJ} is that we have $8\cos^2(\pi k/M)$ instead of $32\cos^2(\pi k/M)$ for the
reasons discussed in the previous subsection.) The above discussion then implies 
the following solution to the twisted tadpole cancellation conditions for the $\Omega J$ 
orientifold of Type IIB on ${\cal M}$ with $M=3$ ($N=(n_3+2\eta)/3)$:
\begin{eqnarray}
 \gamma_{1,3}={\mbox{diag}}(
 \exp(2\pi i/3)~(N~{\mbox{times}}),
 \exp(-2\pi i/3)~(N~{\mbox{times}}),
 1~({N-2\eta}~{\mbox{times}}))~.
\end{eqnarray} 
Here $\eta=-1$ if the $\Omega$ projection is of the $SO$ type, and $\eta=+1$ if it is
of the $Sp$ type\footnote{In this case we have D3-branes only and (unlike in the cases 
with even $M$) there is a choice for the action of $\Omega$ on the Chan-paton charges.}. 
The massless spectra (for both choices of $\eta$) of these models are given in Table
\ref{spectrum1}. 

{}Note that the one-loop $\beta$-functions for the non-Abelian subgroups of the 33
open string sector gauge groups vanish in these models\footnote{Just as in the previous 
cases, the Abelian gauge 
couplings run in these models so that $U(1)$'s decouple at low energies.}. Thus, we 
have ${\cal N}=2$ superconformal gauge theories living in the world-volumes
of the D3-branes.

\subsection{Comments}

{}In this subsection we would like to comment on some of the issues relevant for
the ${\cal N}=2$ superconformal models constructed in the previous subsections.

{}First, note that the world-sheet parity reversal $\Omega$ maps the $g^k$ twisted 
closed string sector to the $g^{M-k}$ twisted closed string sector. This implies that 
(for $k\not=0,M/2$) the $g^k$ and $g^{M-k}$ twisted closed string sectors together
give rise to one hypermultiplet and one vector multiplet. On the other hand, the 
${\bf Z}_2$ twisted sector (present for $M\in2{\bf N}$) gives rise to one hypermultiplet
only.

{}The second comment concerns finiteness of ${\bf Z}_6$ and ${\bf Z}_3$ theories. 
From our discussion in section \ref{finite1} we (at least naively) expect that models
where twisted Chan-Paton matrices are not traceless should not be conformal. Note 
that in the ${\bf Z}_2$ and ${\bf Z}_4$ models discussed above the twisted Chan-Paton
matrices are traceless so it is not surprising that they are finite. On the other hand, some 
of the twisted Chan-Paton matrices in the ${\bf Z}_6$ and ${\bf Z}_3$ models are {\em 
not} traceless. Thus, it might appear surprising that they are still finite. Apparently, some 
subtle cancellation has taken place in these models\footnote{This cancellation need 
occur at one-loop level only since these models have ${\cal N}=2$ supersymmetry.}.

{}We can understand the origin of this ``accidental'' cancellation in the light of recent
results obtained in \cite{NS}. In \cite{NS} it was shown that the six dimensional 
orientifold of Type IIB on $T^4/{\bf Z}_6$ is on the same moduli as the orientifold
of Type IIB on $T^4/{\bf Z}_2$ with an (untwisted) NS-NS antisymmetric tensor 
background. The twisted sector Chan-Paton matrices in the ${\bf Z}_2$ model
are traceless (for turning on the untwisted NS-NS $B$-field does not affect the
twisted tadpole cancellation conditions \cite{NS}). Since the four dimensional 
orientifolds we are considering here are related to the six dimensional ones (upon
compactifying on $T^2$, T-dualizing and taking the dimensions of the dual 
${\widetilde T}^2$ to infinity), it is not surprising that the ${\bf Z}_6$ model is finite
(that is, this explains the ``accidental'' cancellation mentioned above). The finiteness
of the ${\bf Z}_3$ model then follows as ${\bf Z}_6\approx{\bf Z}_3\otimes{\bf Z}_2$,
and the ${\bf Z}_2$ model is finite due to tracelessness of the twisted Chan-Paton 
matrices.    

{}Here we should mention that there are other ${\cal N}=2$ models which are not
superconformal. One way of constructing such models is to consider the $\Omega
{\widetilde J}$ orientifold of Type IIB on ${\cal M}={\bf C}^3/\Gamma$, where 
$\Gamma\approx Z_M$ has odd order ($M\in 2{\bf N}+1$). Note that in these models
there is an orientifold 7-plane (which requires presence of 8 D7-branes) but no orientifold
3-plane\footnote{These models are ``T-dual'' (in the non-compact limit) to some of the six dimensional models discussed in \cite{BI}.}. 
Nonetheless, we can introduce an arbitrary number of D3-branes. The gauge group
coming from the D3-branes is (generically) a product of unitary subgroups, and contains
bi-fundamental matter just as in \cite{LNV,BKV}. However, there is additional matter
coming from the 37 open string sector (which gives fundamentals in the 33 gauge group
which also transform as fundamentals under the 77 gauge group). The twisted
tadpole cancellation
conditions are given by (\ref{odd}). It is not difficult to show that in these models (the
non-Abelian part of) the 33 gauge theory is not superconformal (and, in particular, the
corresponding one-loop $\beta$-functions are non-vanishing). Note that this is in accord 
with expectations in section \ref{finite1}. (The ``accidental'' finiteness of the gauge theories 
discussed in the previous subsections is therefore a special feature of those models.)
It would be interesting to study these non-finite gauge theories further 
in the present string theory framework for they may bring additional insight relevant for
understanding more general non-conformal gauge theories.

\section{${\cal N}=1$ Gauge Theories}\label{N1} 

{}In this section we construct a four dimensional ${\cal N}=1$ gauge 
theory which has vanishing $\beta$-function to all orders in perturbation theory 
in the large $N$ limit. 
This particular gauge theory is obtained by appropriately orientifolding Type IIB on
${\bf C}^3/({\bf Z}_2\otimes {\bf Z}_2)$. We will also discuss orientifolds
corresponding to other orbifold groups (which are subgroups of $SU(3)$) that lead to 
${\cal N}=1$ gauge theories which are {\em not} superconformal.

\subsection{The ${\bf Z}_2\otimes {\bf Z}_2$ Model}

{}We start from Type IIB string theory on ${\cal M}$=
${\bf C}^3/\Gamma$, where $\Gamma=\{1,R_1,R_2,R_3\}\approx {\bf Z}_2
\otimes {\bf Z}_2$ ($R_i R_j=R_k$, $i\not=j\not=k\not=i$) is the
orbifold group whose action on the complex coordinates $z_i$ ($i=1,2,3$) 
on ${\bf C}^3$ is given by $R_i z_j=-(-1)^{\delta_{ij}} z_j$. 
Next, we consider an orientifold of this theory where the 
orientifold action is given by $\Omega J$. The orientifold group is given by
${\cal O}=\{1,R_1,R_2,R_3,\Omega J, \Omega R_1,\Omega R_2, \Omega R_3\}$.

{}This model is a ``T-dual'' (in the non-compact limit) of the model studied in \cite{BL}.
The untwisted tadpole cancellation conditions require presence of three sets of 
D7-branes with 8 D7-branes in each set. Thus, the locations of D$7_i$-branes are 
given by points in the $z_i$ complex plane. The number of D3-branes is unconstrained 
(which is due to the fact that the space transverse to the D3-branes is non-compact). The twisted
tadpole cancellation conditions imply that the corresponding Chan-Paton matrices
$\gamma_{R_i,3}$ and $\gamma_{R_i,7_j}$ are traceless:
\begin{equation}
 {\mbox{Tr}}(\gamma_{R_i,3})={\mbox{Tr}}(\gamma_{R_i,7_j})=0~.
\end{equation} 
A choice\footnote{This choice is unique up to equivalent representations \cite{BL}.} 
consistent
with requirements that the Chan-Paton matrices form a (projective) representation
of (the double cover) of $\Gamma$ is given by ($N=n_3/2$)
\begin{eqnarray}
 \gamma_{R_i,3}=i\sigma_i\otimes {\bf I}_N~,
\end{eqnarray} 
where $\sigma_i$ are Pauli matrices, and ${\bf I}_N$ is an $N\times N$ identity matrix.
(The action on the D$7_i$ Chan-Paton charges
is similar.) The spectrum of this model is given in Table \ref{spectrum2}.

{}Note that the one-loop $\beta$-function for the 33
open string sector gauge group vanishes in this 
model\footnote{For $N>4$ the $7_i 7_i$ gauge groups are 
infrared free. They can therefore be treated as global symmetries (in the context
of the four dimensional gauge theory living on the D3-branes) at low energies.}.
Moreover, all the twisted Chan-Paton matrices are traceless in this model. Following 
our discussion in section \ref{finite1} we thus conclude that this four 
dimensional gauge theory is superconformal (to all loop orders) 
in the large $N$ limit.

{}For completeness, let us give the tree-level superpotential for this model. Let
$\Phi_i$, $\Phi^i_j$, $Q^i$ and $Q^{ij}$ be the matter fields in the 33, $7_i 7_i$,
$37_i$ and $7_i7_j$ open string sectors, respectively. The subscript in 
$\Phi_i$ and $\Phi^i_j$ labels three different chiral superfields (see Table 
\ref{spectrum2}) in the 33 and $7_i 7_i$ sectors. The superpotential can be
computed as in \cite{BL} and is given by (here we suppress the actual
values of the Yukawa couplings)
\begin{equation}
 {\cal W}=\epsilon_{ijk}\Phi_i\Phi_j\Phi_k + 
 \epsilon_{ijk}\Phi^l_i\Phi^l_j\Phi^l_k +
 \epsilon_{ijk}\Phi^i_k Q^{ij} Q^{ij} +
 \Phi_i Q^i Q^i+
 Q^{ij} Q^{jk} Q^{ki} +
 Q^{ij} Q^i Q^j~.
\end{equation}
Here the summation over repeated indices is understood. 

\subsection{Other ${\cal N}=1$ Gauge Theories from Orientifolds}

{}In this subsection we consider ${\cal N}=1$ orientifolds with orbifold groups
such that twisted Chan-Paton matrices are {\em not} traceless. Unlike the ${\cal N}=2$
examples (where certain ``accidental'' cancellations at one-loop order led to finiteness
as discussed in the previous section) the ${\cal N}=1$ models no longer have vanishing
one-loop $\beta$-functions (as expected from our discussion in section \ref{finite1}).

{}Instead of being most general, for illustrative purposes in this subsection we will confine
our attention to $\Omega J$ orientifolds of Type IIB on ${\bf C}^3/\Gamma$ where the
orbifold group $\Gamma=\{g^k\vert k=0,\dots,M-1\}\approx {\bf Z}_M$ ($M$ is odd)
is a subgroup of $SU(3)$ (but not of $SU(2)$). The action of $\Gamma$ on the complex 
coordinates $z_i$ ($i=1,2,3$) on ${\bf C}^3$ is given by $gz_i=\omega^{\ell_i} z_i$, 
where $\omega=\exp(2\pi i/M)$, $\ell_i\not=0$, and $\ell_1+\ell_2+\ell_3=M$.

{}In these orientifolds we have D3-branes only whose number is arbitrary. The twisted 
tadpole cancellation conditions for these orientifolds are isomorphic (upon interchanging
the corresponding D3- and D9-brane Chan-Paton matrices) 
to those for the $\Omega$ orientifolds of Type IIB on ${\bf C}^3/\Gamma$. The latter  
tadpole cancellation conditions were derived in \cite{KS}. Applying those results to
the cases under consideration we have the following twisted tadpole cancellation 
conditions ($k=1,\dots,N-1$):
\begin{equation}\label{tadpoles1}
 {\mbox{Tr}}(\gamma_{2k,3})= -4\eta\prod_{i=1}^3 (1+\omega^{k\ell_i})~.
\end{equation}    
Here $\eta=-1$ if the $\Omega$ projection is of the $SO$ type, and $\eta=+1$ if it is of the
$Sp$ type.

{}It is not difficult to solve these tadpole cancellation conditions for the general case. 
For illustrative purposes, however, here we will discuss only two particular examples.

{}First consider the case where $M=3$. This model is a ``T-dual'' (in the non-compact limit)
of the model studied in \cite{Sagnotti}\footnote{Also see, {\em e.g.}, \cite{LPT}.}. 
The solution to the twisted tadpole cancellation
conditions reads ($N=(n_3-4\eta)/3$):
\begin{eqnarray}
 \gamma_{1,3}={\mbox{diag}}(
 \exp(2\pi i/3)~(N~{\mbox{times}}),
 \exp(-2\pi i/3)~(N~{\mbox{times}}),
 1~({N+4\eta}~{\mbox{times}}))~.
\end{eqnarray}
The massless spectra (for both choices of $\eta$) of these models are given in
Table \ref{spectrum3}. The non-Abelian gauge anomaly is cancelled in this model.
However, there is an anomalous $U(1)$. (The superpotential for this model can be found
in \cite{Sagnotti,ZK}.)

{}Next, let us consider the case where $M=7$. This model is a ``T-dual'' (in the non-compact limit) of the model studied in \cite{KS}.
The solution to the twisted tadpole cancellation
conditions reads ($N=(n_3+4\eta)/7$):
\begin{eqnarray}
 \gamma_{1,3}={\mbox{diag}}(&&
 \exp(2\pi i/7)~(N~{\mbox{times}}),
 \exp(-2\pi i/7)~(N~{\mbox{times}}),\nonumber\\
 &&\exp(4\pi i/7)~(N~{\mbox{times}}),
 \exp(-4\pi i/7)~(N~{\mbox{times}}),\nonumber\\
 &&\exp(6\pi i/7)~(N~{\mbox{times}}),
 \exp(-6\pi i/7)~(N~{\mbox{times}}),1~({N-4\eta}~{\mbox{times}}))~.
\end{eqnarray}
The massless spectra (for both choices of $\eta$) of these models are given in
Table \ref{spectrum3}. The non-Abelian gauge anomaly is cancelled in this model.
However, there is an anomalous $U(1)$. (The superpotential for this model can be found
in \cite{KS}.)

{}Note that none of the above two models have vanishing one-loop $\beta$-functions
for the non-Abelian factors in the gauge group. Moreover, the presence of anomalous
$U(1)$ \cite{anom} implies that there is a Fayet-Iliopoulos D-term which must be cancelled 
via a generalized Green-Schwarz mechanism \cite{GS}. Thus, some fields must acquire 
non-zero vevs to cancel the Fayet-Iliopoulos D-term. Using Type I-Heterotic duality 
\cite{PW} it was argued in \cite{ZK} that the twisted closed string sector states transform
non-trivially under the anomalous $U(1)$ gauge transformation, and (together with the 
dilaton plus axion supermultiplet) can cancel the Fayet-Iliopoulos D-term. In this process
a non-zero vev is generated for the field responsible for this cancellation.

{}Thus, in the ${\cal N}=1$ cases we see that the models with non-vanishing traces of
the twisted Chan-Paton matrices lead (as expected in section \ref{finite1}) to non-finite
models. In particular, the ``accidental'' vanishing of the one-loop $\beta$-functions (as in
the corresponding ${\cal N}=2$ cases) does not occur.

{}Here we should mention that the tadpole cancellation condition (\ref{tadpoles1}) 
is necessary not only for ultraviolet finiteness of the corresponding theories but also
for non-Abelian gauge anomaly cancellation. For illustrative purposes, to see what 
can go wrong if we relax this condition, let us consider the above ${\bf Z}_3$ example
with the following choice for the twisted Chan-Paton matrices:
\begin{eqnarray}
 \gamma_{1,3}={\mbox{diag}}(
 \exp(2\pi i/3)~(N~{\mbox{times}}),
 \exp(-2\pi i/3)~(N~{\mbox{times}}),
 1~(N^\prime~{\mbox{times}}))~.
\end{eqnarray}  
The gauge group in the 33 open string sector is $U(N)\otimes G_\eta (N^\prime)$, and
the chiral matter is given by $3({\bf R}_\eta,{\bf 1})$ and $3({\overline {\bf N}}, {\bf N}^\prime)$
(see Table \ref{spectrum3} for notation). Note that ${\bf R}_\eta$ of $SU(N)$ contributes
as much as $N+4\eta$ chiral superfields in ${\bf N}$ of $SU(N)$ into the non-Abelian gauge
anomaly. Thus, the non-Abelian gauge anomaly cancellation implies that 
$N^\prime=N+4\eta$. This is precisely the solution to the tadpole cancellation condition 
(\ref{tadpoles1}).

\subsection{Cases without World-Sheet Description}

{}In subsection C of section \ref{finite1} we mentioned that for certain choices
of the orbifold group the corresponding orientifolds may not have a world-sheet
description. Here we give an explicit example of this.

{}Consider Type IIB on ${\bf C}^3/\Gamma$ with $\Gamma\approx D_N$ (non-Abelian
dihedral group) where the action of $\Gamma$ on the complex coordinates $z_i (i=1,2,3)$
on ${\bf C}^3$ is given by ($\omega=\exp(2\pi i/N)$):
\begin{eqnarray}
 &&gz_1=z_1~,~~~gz_2=\omega z_2~,~~~gz_3=\omega^{-1} z_3~,\\
 &&rz_1=-z_1~,~~~rz_2=z_3~,~~~rz_3=z_2~,
\end{eqnarray}
where $g,r$ are the generators of $D_N$. Note that $g$ and $r$ do not commute:
$rg=g^{-1}r$.

{}Now consider $\Omega J$ ($Jz_i=-z_i$) orientifold of this theory. 
The orientifold group is ${\cal O}=\{g^k,rg^k,\Omega Jg^k,\Omega Jrg^k\vert
k=0,\dots,N-1\}$. Note that $(Jrg^k)^2=1$, and
the set of points in ${\bf C}^3$ fixed under the action of $Jrg^k$ has real dimension 
two. This implies that there are $N$ kinds of orientifold 7-planes corresponding to the
elements $\Omega Jrg^k$. Note, however, that due to non-commutativity between 
$g$ and $r$ (and, therefore, between different $Jrg^k$), these orientifold 7-planes
(as well as the corresponding D7-branes) are mutually non-local. This implies that
this orientifold does not have a world-sheet description. One way to understand such
models (at least in compact cases) is to use their F-theory \cite{vafa} description.

{}Here we should mention that there are other models with both Abelian and non-Abelian
orbifold groups for which the perturbative description in terms of open strings stretched 
between mutually local D-branes is inadequate. We refer the reader to \cite{KST} for details.

\acknowledgments

{}I would like to thank Michael Bershadsky, Andrei Johansen, Henry
Tye, and especially Cumrun Vafa for valuable discussions.
This works is supported in part by the grant
NSF PHY-96-02074, and the DOE 1994 OJI award. I would also like to
thank Albert and Ribena Yu for financial support.

\newpage
\begin{figure}[t]
\epsfxsize=16 cm
\epsfbox{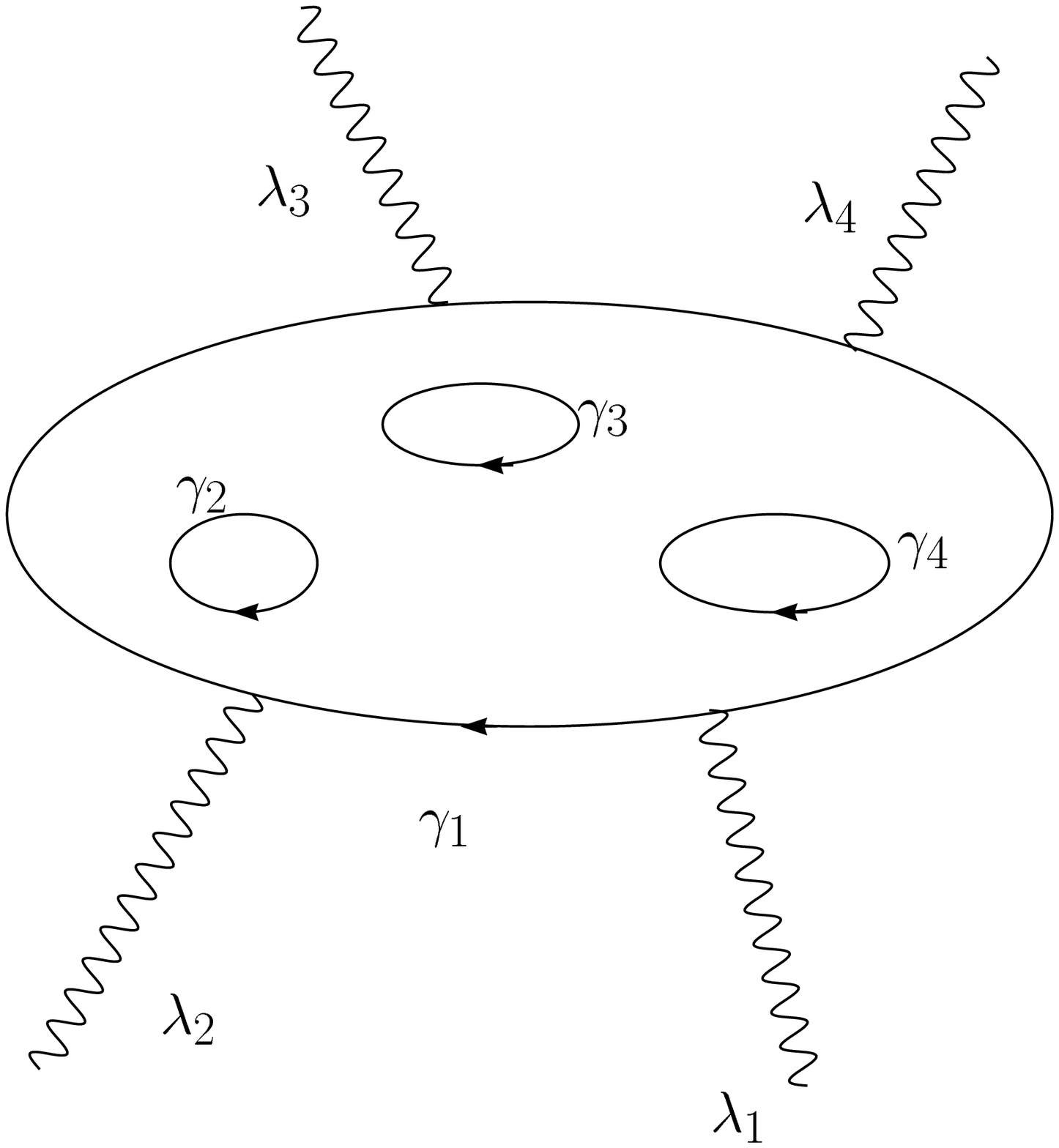}
\caption{A planar diagram.}
\end{figure}

\newpage
\begin{figure}[t]
\hspace*{3 cm}
\epsfxsize=10 cm
\epsfbox{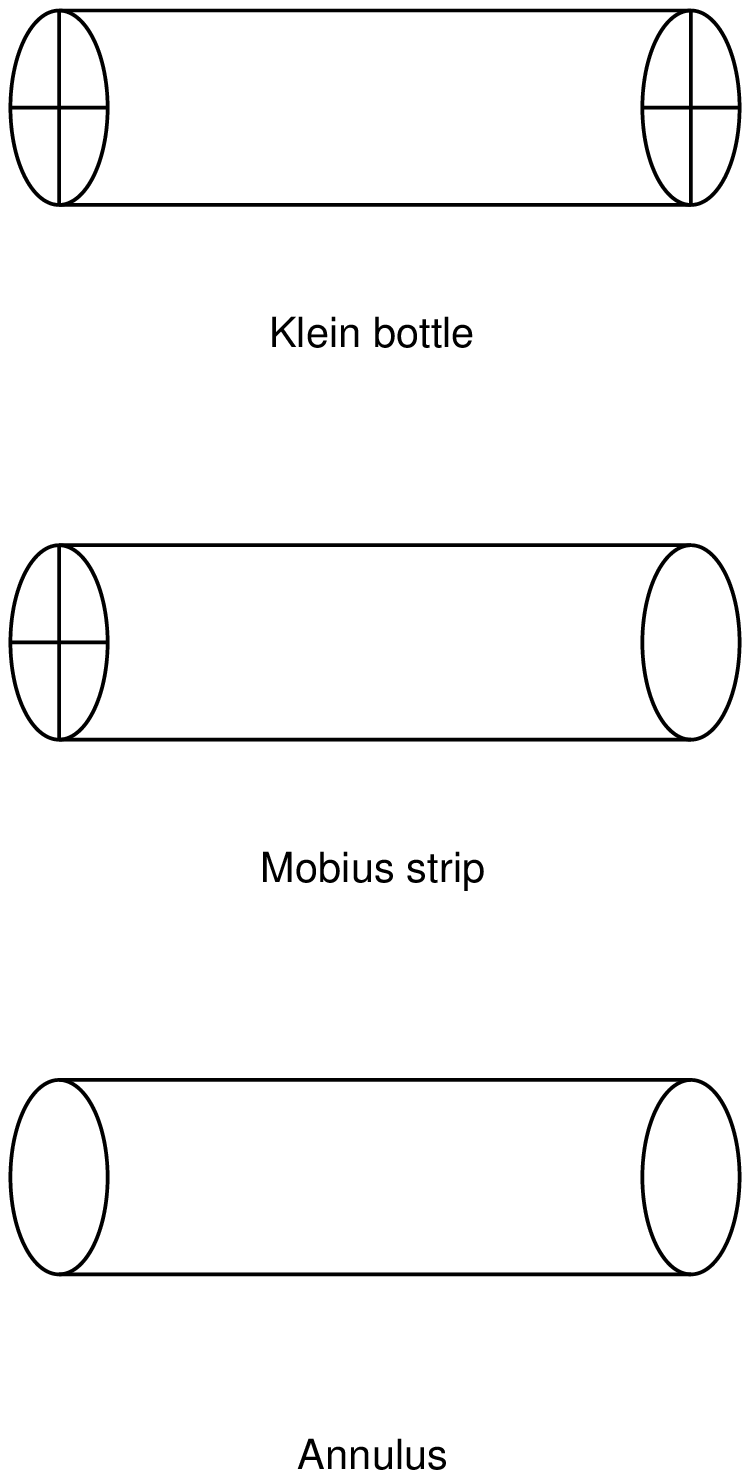}
\caption{Tree-channel Klein bottle, M{\"o}bius strip and annulus amplitudes.}
\end{figure}

\newpage
\begin{figure}[t]
\hspace*{3 cm}
\epsfxsize=10 cm
\epsfbox{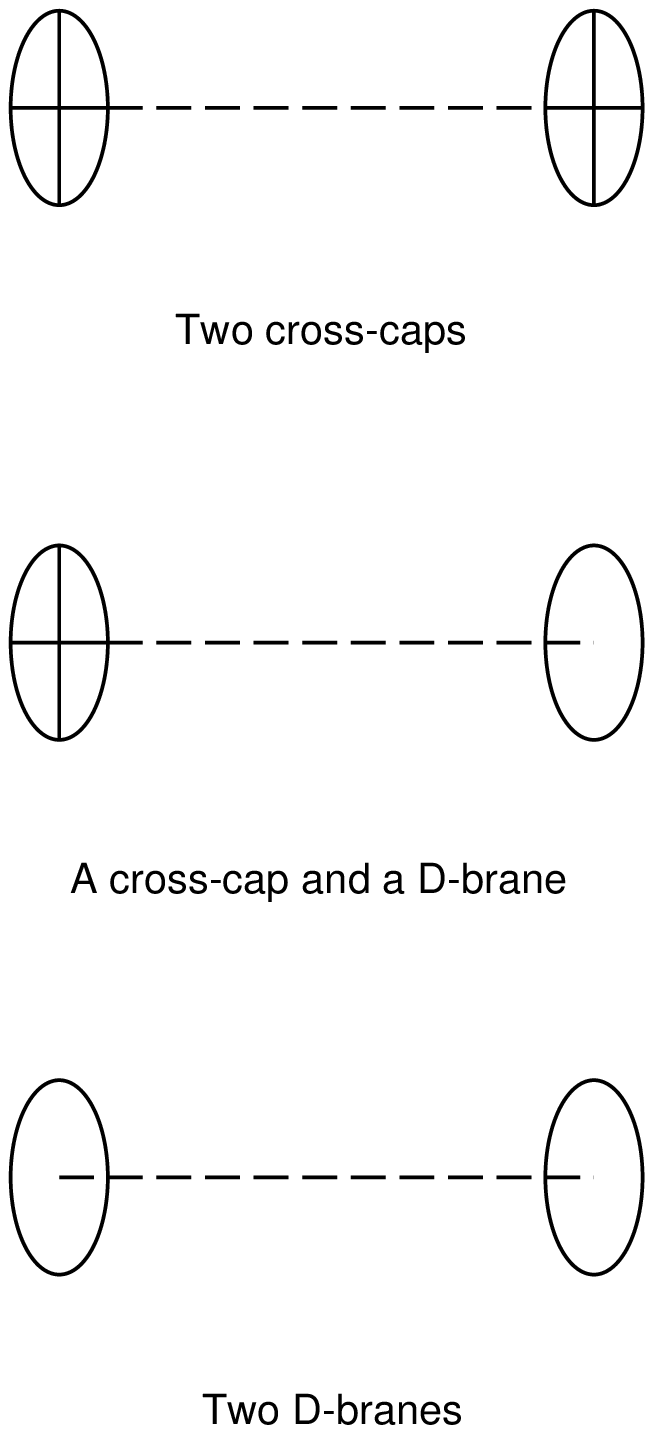}
\caption{Factorization of the Klein bottle, M{\"o}bius strip and annulus amplitudes.}
\end{figure}

\begin{table}[t]
\begin{tabular}{|c|c|l|c|c|}
 Model & Gauge Group & \phantom{Hy} Charged  & Twisted Sector 
& Twisted Sector  \\
       &                &Hypermultiplets & Hypermultiplets
&Vector Multiplets \\
\hline
${\bf Z}_2$ & $U(N)_{33} \otimes U(4)_{77}$ & 
 $2 ({\bf A};{\bf 1})_{33}$ & $1$
& $0$ \\
               &                       & $2 ({\bf 1};{\bf 6})_{77}$ & & \\
               &                       & $({\bf N};{\bf 4})_{37}$ & & \\
\hline
${\bf Z}_4$  & $[U(N) \otimes U(N)]_{33}\otimes$ 
& $({\bf A},{\bf 1};{\bf 1},{\bf 1})_{33}$ & $2$ & $1$ \\ 
 &$[U(2) \otimes U(2)]_{77}$ & $({\bf 1},{\bf A};{\bf 1},{\bf 1})_{33}$ & & \\
 & & $({\bf N},{\bf N};{\bf 1},{\bf 1})_{33}$ & & \\
 & & $({\bf 1},{\bf 1};{\bf 1}^\prime,{\bf 1})_{77}$ & & \\
 & & $({\bf 1},{\bf 1};{\bf 1},{\bf 1}^\prime)_{77}$ & & \\
 & & $({\bf 1},{\bf 1};{\bf 2},{\bf 2})_{77}$ & & \\
 & & $({\bf N},{\bf 1};{\bf 2},{\bf 1})_{37}$ & & \\
 & & $({\bf 1},{\bf N};{\bf 1},{\bf 2})_{37}$ & & \\
\hline
${\bf Z}_6$  & $[U(N) \otimes U(N) \otimes U(N+1)]_{33}\otimes$ 
&$({\bf A},{\bf 1},{\bf 1};{\bf 1})_{33}$ & $3$ & $2$ \\
 & $[U(1) \otimes U(1) \otimes U(2)]_{77}$&
$({\bf 1},{\bf A},{\bf 1};{\bf 1})_{33}$  & & \\
 & & $({\bf N},{\bf 1},{\bf N+1};{\bf 1})_{33}$ & & \\
 & & $({\bf 1},{\bf N},{\bf N+1};{\bf 1})_{33}$ & & \\
 & & $({\bf 1},{\bf 1},{\bf 1};{\bf 2}_1)_{77}$ & & \\
 & & $({\bf 1},{\bf 1},{\bf 1};{\bf 2}_2)_{77}$ & & \\
 & & $({\bf N},{\bf 1},{\bf 1};{\bf 1}_1)_{37}$ & & \\
 & & $({\bf 1},{\bf N},{\bf 1};{\bf 1}_2)_{37}$ & & \\
 & & $({\bf 1},{\bf 1},{\bf N+1};{\bf 2})_{37}$ & & \\
\hline
${\bf Z}_3$ & $[U(N) \otimes G_\eta (N-2\eta)]_{33}$ & $({\bf R}_\eta,{\bf 1})_{33}$ & $1$
& $1$ \\
 & & $({\bf N},{\bf N-2\eta})_{33}$ & & \\
\hline
\end{tabular}
\caption{The massless spectra of ${\cal N}=2$ orientifolds of Type IIB on ${\bf C}^3/{\bf Z}_N$
$M=2,3,4,6$. The semi-colon in the column ``Charged Hypermultiplets'' separates $33$ and 
$77$ representations. The notation ${\bf A}$ stands for the two-index antisymmetric
representation (which is $N(N-1)/2$ dimensional for $U(N)$) of the corresponding unitary 
group. The notation ${\bf 1}^\prime$ in the ${\bf Z}_4$ model stands for the antisymmetric 
representation of the corresponding $U(2)$ group (whose charge with respect to the $U(1)$
factor is 2). In the ${\bf Z}_6$ model the subscripts ``1'' and ``2'' indicate that the corresponding
states are charged under the first and the second $U(1)$ factors in the 77 gauge group. 
In the ${\bf Z}_3$ model $G_\eta=SO$ for $\eta=-1$ and $G_\eta=Sp$ for $\eta=+1$. (Here 
we are using the convention that $Sp(2m)$ has rank $m$.) Also, $R_\eta={\bf A}$ for $\eta=-1$,
and $R_\eta={\bf S}$ (two-index $N(N+1)/2$ dimensional symmetric representation of $U(N)$)
for $\eta=+1$.
By twisted sector hypermultiplets/vector multiplets we mean those in the twisted {\em closed} 
string sectors. The untwisted closed string sector states are not shown.}
\label{spectrum1} 
\end{table}

\begin{table}[t]
\begin{tabular}{|c|c|l|c|c|}
 Model & Gauge Group & \phantom{Hy} Charged  & Twisted Sector 
& Twisted Sector  \\
       &                &Chiral Multiplets & Chiral Multiplets
&Vector Multiplets \\
\hline
${\bf Z}_2\otimes{\bf Z}_2$ & $Sp(N)_{33} \otimes$  & 
 $3 ({\bf A};{\bf 1};{\bf 1};{\bf 1})_{33}$ & $3$
& $0$ \\
               &  $Sp(4)_{7_1 7_1}\otimes Sp(4)_{7_2 7_2}\otimes Sp(4)_{7_3 7_3}$   
                  & $3 ({\bf 1};{\bf 6};{\bf 1};{\bf 1})_{7_1 7_1}$ & & \\
                &  & $3 ({\bf 1};{\bf 1};{\bf 6};{\bf 1})_{7_2 7_2}$ & & \\
                &  & $3 ({\bf 1};{\bf 1};{\bf 1};{\bf 6})_{7_3 7_3}$ & & \\
                &  & $({\bf N};{\bf 4};{\bf 1};{\bf 1})_{3 7_1}$ & & \\
                &  & $({\bf N};{\bf 1};{\bf 4};{\bf 1})_{3 7_2}$ & & \\
                &  & $({\bf N};{\bf 1};{\bf 1};{\bf 4})_{3 7_3}$ & & \\
                &  & $({\bf 1};{\bf 4};{\bf 4};{\bf 1})_{7_1 7_2}$ & & \\
                &  & $({\bf 1};{\bf 1};{\bf 4};{\bf 4})_{7_2 7_3}$ & & \\
                &  & $({\bf 1};{\bf 4};{\bf 1};{\bf 4})_{7_3 7_1}$ & & \\
\hline
\end{tabular}
\caption{The massless spectrum of ${\cal N}=1$ orientifold of Type IIB on ${\bf C}^3/
({\bf Z}_2\otimes {\bf Z}_2)$.
The notation ${\bf A}$ stands for the two-index antisymmetric (reducible)
representation of $Sp(N)$. The untwisted closed string sector states are not shown.}
\label{spectrum2} 
\end{table}

\begin{table}[t]
\begin{tabular}{|c|c|l|c|c|}
 Model & Gauge Group & \phantom{Hy} Charged  & Twisted Sector 
& Twisted Sector  \\
       &                &Chiral Multiplets & Chiral Multiplets
&Vector Multiplets \\
\hline
${\bf Z}_3$ & $[U(N)\otimes G_\eta(N+4\eta)]_{33}$  & 
 $3 ({\bf R}_\eta,{\bf 1})(+2)_{33}$ & $1$
& $0$ \\
                &  & $3({\overline {\bf N}},{\bf N+4\eta})(-1)_{33}$ & & \\
\hline
${\bf Z}_7$ & $[U(N)\otimes U(N)\otimes U(N)\otimes$  & 
 $({\bf R}_\eta,{\bf 1},{\bf 1},{\bf 1})(+2,0,0)_{33}$ & $3$
 & $0$ \\
            &  $G_\eta(N-4\eta)]_{33}$ & $({\bf 1},{\bf R}_\eta,{\bf 1},{\bf 1})(0,+2,0)_{33}$ & & \\
            &  & $({\bf 1},{\bf 1},{\bf R}_\eta,{\bf 1})(0,0,+2)_{33}$ & & \\
                &  & $({\bf N},{\bf 1},{\bf 1},{\bf N-4\eta})(+1,0,0)_{33}$ & & \\
                &  & $({\bf 1},{\bf N},{\bf 1},{\bf N-4\eta})(0,+1,0)_{33}$ & & \\
               &  & $({\bf 1},{\bf 1},{\bf N},{\bf N-4\eta})(0,0,+1)_{33}$ & & \\
               &  & $({\overline {\bf N}},{\bf N},{\bf 1},{\bf 1})(-1,+1,0)_{33}$ & & \\
               &  & $({\bf 1},{\overline {\bf N}},{\bf N},{\bf 1})(0,-1,+1)_{33}$ & & \\
               &  & $({\bf N},{\bf 1},{\overline {\bf N}},{\bf 1})(+1,0,-1)_{33}$ & & \\
              &  & $({\overline {\bf N}},{\overline {\bf N}},{\bf 1},{\bf 1})(-1,-1,0)_{33}$ & & \\
               &  & $({\bf 1},{\overline {\bf N}},{\overline {\bf N}},{\bf 1})(0,-1,-1)_{33}$ & & \\
               &  & $({\overline {\bf N}},{\bf 1},{\overline {\bf N}},{\bf 1})(-1,0,-1)_{33}$ & & \\
\hline
\end{tabular}
\caption{The massless spectra of ${\cal N}=1$ orientifolds of Type IIB on ${\bf C}^3/
{\bf Z}_M$ ($M=3,7$). Here we are using some of the notations from Table I.
The $U(1)$ charges of the states in the 33 open string sector are
given in parentheses. The untwisted closed string sector states are not shown.}
\label{spectrum3} 
\end{table}


\begin{references}


\bibitem{BKV} M. Bershadsky, Z. Kakushadze and C. Vafa, 
``String Expansion as Large $N$ Expansion of Gauge Theories'', hep-th/9803076.

\bibitem{thooft} G. `t Hooft, ``A Planar Diagram Theory For Strong
Interactions'', Nucl. Phys. {\bf B72} (1974) 461.

\bibitem{CS} E. Witten, ``Chern-Simons Gauge Theory As A String Theory'',
hep-th/9207094. 

\bibitem{KaSi} S. Kachru and E. Silverstein, ``4d Conformal Field Theories
and Strings on Orbifolds'', hep-th/9802183.

\bibitem{LNV} A. Lawrence, N. Nekrasov and C. Vafa, ``On Conformal Theories in 
Four Dimensions'', hep-th/9803015.

\bibitem{Iban} L.E. Ib{\'a}{\~n}ez, ``A Chiral D=4, N=1 String Vacuum with a 
Finite Low Energy Effective Field Theory'', hep-th/9802103. 

\bibitem{HSU} A. Hanany, M.J. Strassler and A.M. Uranga, ``Finite Theories and 
Marginal Operators on the Brane'', hep-th/9803086.

\bibitem{Koso} D.A. Kosower, B.-H. Lee and V.P. Nair, ``Multi Gluon Scattering:
A String Based Calculation'', Phys. Lett. {\bf B201} (1988) 85;\\
Z. Bern and D.A. Kosower, ``Efficient Calculation of One Loop QCD Amplitudes'',
Phys. Rev. Lett. {\bf 66} (1991) 1669.
\bibitem{Dix} Z. Bern, L. Dixon, D.C. Dunbar, M. Perelstein and J.S. Rozowsky, 
``On the Relationship between Yang-Mills Theory and Gravity and Its Implication for
Ultraviolet Divergences'', hep-th/9802162.

\bibitem{Kleb} I.R. Klebanov, ``World Volume Approach to Absorption by
Non-dilatonic Branes'', Nucl. Phys. {\bf B496} (1997) 231, hep-th/9702076.

\bibitem{Gubs} S.S. Gubser and I.R. Klebanov, ``Absorption by Branes and Schwinger 
Terms in the World Volume Theory'', Phys. Lett. {\bf B413} (1997) 41,
hep-th/9708005.

\bibitem{Mald} J.M. Maldacena, ``The Large $N$ Limit of
Superconformal Field
Theories and Supergravity'', hep-th/9711200.

\bibitem{Poly} S.S.  Gubser, I.R. Klebanov and A.M. Polyakov, ``Gauge Theory 
Correlators from
Non-Critical String Theory'', hep-th/9802109.

\bibitem{Oogu} G.T. Horowitz and H. Ooguri, ``Spectrum of Large $N$ Gauge
Theory from Supergravity'', hep-th/9802116.

\bibitem{Witt} E. Witten, ``Anti-de Sitter
Space And Holography'', hep-th/9802150; 
``Anti-de Sitter Space, Thermal Phase Transition, And Confinement 
In Gauge Theories'', hep-th/9803131.

\bibitem{related} 
S.S. Gubser, I.R. Klebanov and A.W. Peet, ``Entropy and
Temperature of Black 3-branes'', Phys. Rev. {\bf D54} (1996) 3915,
hep-th/9602135;\\
M.R. Douglas, J. Polchinski and A. Strominger, ``Probing Five Dimensional 
Black Holes with D-branes'', hep-th9703031;\\
S.S. Gubser, I.R. Klebanov and A.A. Tseytlin, ``String Theory and Classical
Absorption by Three Branes'', Nucl. Phys. {\bf B499} (1997) 217, hep-th/9703040;\\
J.M. Maldacena and A. Strominger, ``Semiclassical Decay of Near Extremal
Fivebranes'', hep-th/9710014;\\
A.M. Polyakov, ``String Theory and Quark
Confinement'', hep-th/9711002;\\
K. Sfetsos and K. Skenderis, ``Microscopic Derivation of the Bekenstein-Hawking 
Entropy Formula for Non-extremal Black Holes'', hep-th/9711138;\\ 
S. Ferrara and C. Fronsdal, ``Conformal Maxwell Theory as a Singleton
Field Theory on $ADS_5$, IIB Three Branes and Duality'', hep-th/9712239;\\
N. Itzhaki, J.M. Maldacena, J. Sonnenschein and S. Yankielowicz,
``Supergravity and the Large $N$ Limit of Theories With Sixteen Supercharges'',
hep-th/9802042;\\
M. Gunaydin and D. Minic, ``Singletons, Doubletons and M Theory'', hep-th/9802047;\\
M. Berkooz, ``A Supergravity Dual of a (1,0) Field Theory in Six Dimensions'',
hep-th/9802195;\\
V. Balasubramanian and F. Larsen, ``Near Horizon Geometry
and Black Holes
in Four Dimensions'', hep-th/9802198;\\
S. Ferrara, C. Fronsdal and A. Zaffaroni,
``On N=8 Supergravity on $AdS_5$ and N=4 Superconformal Yang-Mills theory'',
hep-th/9802203;\\
S.-J. Rey and J.-T. Yee, ``Macroscopic
Strings as Heavy Quarks of Large $N$ Gauge
Theory and Anti-de Sitter Supergravity'', hep-th/9803001;\\
J.M. Maldacena, ``Wilson loops in large $N$ field theories'',
hep-th/9803002;\\
M. Flato and C. Fronsdal, ``Intersecting Singletons'', hep-th/9803013;\\
S.S.  Gubser, A. Hashimoto, I.R. Klebanov and M. Krasnitz, ``Scalar Absorption
and the Breaking of the Worldvolume Conformal Invariance'', hep-th/9803023;\\
I.Ya. Aref'eva and I.V. Volovich, ``On Large $N$ Conformal Theories, Field
Theories in
the Anti-de Sitter space and Singletons'', hep-th/9803023;\\
L. Castellani, A. Ceresole, R. D'Auria, S. Ferrara, P. Fr{\'e} and M.
Trigiante,
``$G/H$ M-branes and $AdS_{p+2}$ Geometries'', hep-th/9803039;\\
O. Aharony, Y. Oz and Z. Yin, ``M Theory on $AdS_p\times S^{11-p}$ and
Superconformal Field Theories'', hep-th/9803051;\\
S. Minwala, ``Particles on $AdS_{4/7}$ and Primary Operators on $M_{2/5}$
Brane Worldvolumes'', hep-th/9803053;\\
S. Ferrara and A. Zaffaroni, ``N=1,2 4D Superconformal Field Theories and Supergravity
in $AdS_5$'', hep-th/9803060;\\
M.J. Duff, H. Lu and C.N. Pope, ``$AdS_5\times S^5$ Untwisted'', hep-th/9803061;\\
R.G. Leigh and M. Rozali, ``The Large N Limit of the (2,0) Superconformal Field Theory'',
hep-th/9803068;\\
E. Halyo, ``Supergravity on $AdS_{4/7}\times S^{7,4}$ and M Branes'', hep-th/9803077;
``Supergravity on $AdS_{5/4} \times$ Hopf Fibrations and Conformal Field 
Theories'', hep-th/9803193;\\
A. Rajaraman, ``Two-Form Fields and the Gauge Theory Description of Black Holes'',
hep-th/9803082;\\
G.T. Horowitz and S.F. Ross, ``Possible Resolution of Black Hole Singularities from Large 
N Gauge Theory'', hep-th/9803085;\\
S. Ferrara, A. Kehagias, H. Partouche and A. Zaffaroni, ``Membranes and Fivebranes
with Lower Supersymmetry and their AdS Supergravity Duals'', hep-th/9803109;\\
J.A. Minahan, ``Quark-Monopole Potentials in Large N Super Yang-Mills'', hep-th/9803111;\\
J. Gomis, ``Anti de Sitter Geometry and Strongly Coupled Gauge Theories'', 
hep-th/9803019;\\
S.-J. Rey, S. Theisen and J.-T. Yee, ``Wilson-Polyakov Loop at Finite Temperature in Large 
N Gauge Theory and Anti-de Sitter Supergravity'', hep-th/9803135;\\
A. Brandhuber, N. Itzhaki, J. Sinnenschein and S. Yankielowicz, ``Wilson Loops in the
Large N Limit at Finite Temperature'', hep-th/9803137;\\
M. Gunaydin, ``Unitary Supermultiplets of OSp(1/32,R) and M-theory'', hep-th/ 9803138;\\
Y. Oz and J. Terning, ``Orbifolds of $AdS_5\times S^5$ and $4d$ Conformal Field 
Theories'', hep-th/9803167;\\
L. Andrianopoli and S. Ferrara, ``K-K excitations on $AdS_5 \times S^5$
 as N=4 `primary' superfields'', hep-th/9803171;\\
I.V. Volovich, ``Large N Gauge Theories and Anti-de Sitter Bag Model'', hep-th/9803174;\\
F. Brandt, J. Gomis and J. Simon, ``D-string on Near Horizon Geometries and Infinite 
Conformal Symmetry'', hep-th/9803196.




\bibitem{kap} V. Kaplunovsky,
``One-Loop Threshold Effects in String Unification'', Nucl. Phys. {\bf B307}
(1988) 145;
ERRATA for ``One-Loop Threshold Effects in String Unification'',
Nucl. Phys. {\bf B382} (1992) 436, hep-th/9205068.

\bibitem{KST} Z. Kakushadze, G. Shiu and S.-H.H. Tye, ``Type IIB Orientifolds,
F-theory, Type I Strings on Orbifolds and Type I - Heterotic Duality'', preprint
CLNS 98/1549, HUTP-97/A044, NUB 3168, to appear.

\bibitem{Zwart} G. Zwart, ``Four-dimensional $N=1$ $Z_N \times Z_M$ Orientifolds'', hep-th/9708040.

\bibitem{PS} G. Pradisi and A. Sagnotti, ``Open String Orbifolds'',
Phys. Lett. {\bf B216} (1989) 59;\\
M. Bianchi and A. Sagnotti, ``On the Systematics of Open String Theories'',
Phys. Lett. {\bf B247} (1990) 517; ``Twist Simmetry and Open String Wilson Lines'',
Nucl. Phys. {\bf B361} (1991) 519. 

\bibitem{GP} E.G. Gimon and J. Polchinski, ``Consistency Conditions 
for Orientifolds and D-Manifolds'', Phys. Rev. {\bf D54} (1996) 1667, hep-th/9601038.

\bibitem{GJ}
E.G. Gimon and C.V. Johnson, ``K3 Orientifolds'' Nucl. Phys. {\bf B477} (1996) 715, 
hep-th/9604129;\\
A. Dabholkar and J. Park, ``Strings on Orientifolds'', Nucl. Phys. {\bf B477} (1996) 701, 
hep-th/9604178.

\bibitem{NS} Z. Kakushadze, G. Shiu and S.-H.H. Tye, 
``Type IIB Orientifolds
with NS-NS Antisymmetric Tensor Backgrounds'', hep-th/9803141.

\bibitem{BI} J.D. Blum and K. Intriligator, ``Consistency Conditions for Branes at Orbifold Singularities'', Nucl. Phys. {\bf B506} (1997) 223, hep-th/9705030; 
``New Phases of String Theory and 6d RG Fixed Points via Branes at 
Orbifold Singularities'', Nucl. Phys. {\bf B506} (1997) 199, hep-th/9705044;\\
K. Intriligator, ``New String Theories in Six Dimensions via Branes at Orbifold 
Singularities'', hep-th/9708117.

\bibitem{BL} M. Berkooz and R.G. Leigh, ``A D=4 N=1 Orbifold of Type I Strings'',
Nucl. Phys. {\bf B483} (1997) 187,
hep-th/9605049.

\bibitem{KS} Z. Kakushadze and G. Shiu, ``A Chiral $N=1$ Type I Vacuum in Four Dimensions and Its Heterotic Dual'', Phys. Rev. {\bf D56} (1997) 3686,
hep-th/9705163; ``4D Chiral $N=1$ Chiral Vacua with and 
without $D5$-branes'', 
hep-th/9706051.

\bibitem{Sagnotti} C. Angelantonj, M. Bianchi, G. Pradisi, A. Sagnotti and 
Ya.S. Stanev, ``Chiral Asymmetry in Four-Dimensional Open-String Vacua'', 
Phys. Lett. {\bf B385} (1996) 96, hep-th/9606169.

\bibitem{LPT} J. Lykken, E. Poppitz and S. Trivedi, ``M(ore) on Chiral Gauge Theories
from D-branes'', hep-th/9712193.

\bibitem{PW} J. Polchinski and E. Witten, ``Evidence for Heterotic - Type I String Duality'', Nucl. Phys. {\bf B460} (1996) 525, hep-th/9510169.

\bibitem{ZK} Z. Kakushadze, ``Aspects of $N=1$ Type I-Heterotic Duality in Four 
Dimensions'', Nucl. Phys. {\bf B512} (1998) 221, hep-th/9704059.

\bibitem{anom} E. Witten, ``Some Properties of O(32) Superstrings'', 
Phys. Lett. {\bf B149} (1984) 351;\\
M. Dine, N. Seiberg and E. Witten, ``Fayet-Iliopoulos Terms in String Theory'',
Nucl. Phys. {\bf B289} (1987) 589.

\bibitem{GS} M. Green and J.H. Schwarz, ``Anomaly Cancellations In
Supersymmetric D=10 Gauge Theory Require SO(32)'',
Phys. Lett. {\bf B149} (1984) 117;
``Infinity Cancellations in SO(32) Superstring Theory'', Phys. Lett. {\bf B151} (1985) 21. 

\bibitem{vafa} C. Vafa, ``Evidence for F-theory'', Nucl. Phys. {\bf B469} (1996) 403,
hep-th/9602022.





    


\end{references}
\end{document}